\documentclass[proceedings]{JHEP} % 10pt is ignored!

\def \beq {\begin{equation}} \def \eeq {\end{equation}} \def \lf {\left
(} \def \rt {\right )} \def \a {\alpha} \def \lm {\lambda} \def \D
{\Delta} \def \r {\rho} \def \th {\theta} \def \rg {\sqrt{g}} \def \Slash
{\, / \! \! \! \!}   \def \pl
{\partial}
\def    \del    {\nabla}

\conference{Quantum aspects of gauge theories, supersymmetry and
unification}

\title{Aspects of Holography and Rotating AdS Black Holes}

\author{David Berman \thanks{This is work done in collaboration with
M. Parikh}\\
        Institute for Theoretical Physics\\
         University of Groningen \\
          Nijenborgh 4, 9747 AG Groningen, The Netherlands\\   

        E-mail: \email{d.berman@phys.rug.nl}}

\abstract{ A comparison is made between the thermodynamics of weakly and
strongly coupled Yang-Mills with fixed angular momentum. The free energy
of the strongly coupled Yang-Mills is calculated by using a dual
supergravity description corresponding to a rotating black hole in an
Anti de Sitter (AdS) background. All thermodynamic quantities are shown
have the same ratio of 3/4 (independent of angular momentum) between
strong
and weak coupling. 
}

\begin{document}

\section{Introduction}

The AdS/CFT correspondence \cite{adscft} has provided many insights into
the connection between gauge theories and gravity. (See \cite{adsreview}
for an extensive review of this vast subject.) In some sense the
correspondence provides a duality map between large N Yang-Mills and IIB
supergravity in an $AdS_5 \times S^5$ background. The origin of this
duality relation is from considering the physics of D3 branes from two
perspectives. The world volume description, as can be derived from
considering the D-brane as a submanifold on which strings can end, gives a
Yang-Mills theory. The soliton description, whereby the the D-brane is
seen as a solution in IIB supergravity provides, after taking
appropriate limits, the alternative $AdS$ supergravity point of
view. One may use this duality relation to calculate quantities in the
Yang-Mills theory at large `t Hooft coupling. In particular one may
calculate thermodynamic quantities in the Yang-Mills theory from
considering black holes in the AdS space-time \cite{fourthirds}. Comparing
these thermodynamic quantities with those calculated directly from the 
Yang-Mills theory at weak coupling, one sees that they correspond up to a
factor of four thirds. (It should be stressed that there is no reason to
expect them to match as typically entropies change as the coupling is
altered). 
 
For nonrotating AdS black holes \cite{hawkingpage}, the thermodynamics
has been described by thermal conformal field theory
\cite{holographyads}. Recently a five-dimensional  rotating black hole
embedded in anti-de Sitter space has been discovered \cite{5dadskerr}.
Since rotation introduces an extra dimensionful parameter, the
conformal field theory entropy is not so tightly constrained by the
combination of extensivity and dimensional analysis; thus a
comparison between thermodynamic quantities is much more nontrivial.
Our purpose in this paper, as first reported in \cite{bap}, is to
compare the thermodynamics  of the new rotating black hole with that
of the dual conformal field  theory in four dimensions. If one then
assumes the AdS/CFT correspondence to be correct then the result may
be interpreted as the comparison of thermodynamic quantites in the
field theory at strong and weak coupling.

This correspondence has also been investigated in the case of rotating
D branes by \cite{roberto1,roberto2,steve,steve2,steve3,obers}.

We begin by demonstrating the holographic \cite{hologerard,hololenny}
nature of the duality for nonrotating black holes: the thermodynamics
of a nonrotating black hole in anti-de Sitter space emerges from a
thermal conformal field theory whose thermodynamic variables are read
off from the boundary of the black hole spacetime.  In the high
temperature limit, the field theory calculation gives the correct
entropy of the Hawking-Page black hole up to a factor of 4/3.

We then describe the new rotating Kerr-AdS black hole solution and
show how its thermodynamic properties can be recovered from the dual
field theory, in the high temperature limit.  In that limit, the
entropy, energy and angular momentum, compared with the statistical
mechanics of the field theory, all agree with their gravitational
counterparts, again up to a common factor of 4/3.

\section{AdS/CFT Correspondence for Nonrotating Holes}

The five-dimensional Einstein-Hilbert action with a cosmological
constant is given by  \beq  I = - {1 \over 16 \pi G_5} \int d^5 x
\sqrt{-g} \lf R + 12 l^2 \rt \; ,  \eeq  where $G_5$ is the
five-dimensional Newton constant, $R$ is the Ricci scalar, the
cosmological constant is $\Lambda = - 6 l^2$, and we have neglected a
surface term at infinity.  Anti-de Sitter solutions derived from this
action can be embedded in ten-dimensional IIB supergravity such that
the supergravity background is of the form $AdS_5 \times S^5$. The
AdS/CFT correspondence then states that there  is a dual conformal
field theory in four dimensions from which one can  extract the
physics.

The line element of a ``Schwarzschild'' black hole in anti-de Sitter
space \cite{hawkingpage} in five spacetime dimensions can be written
as  \begin{eqnarray}  ds^2 = - \lf 1 - {2 M G_5 \over r^2} + r^2 l^2
\rt dt^2  \nonumber \\  +  \lf 1 - {2 M G_5 \over r^2} + r^2 l^2
\rt^{\! \! -1} dr^2  + r^2 d \Omega^2_3 \; .  \label{HPds2}
\end{eqnarray} This solution has a horizon at $r = r_+$ where  \beq
r_+^2 = {1 \over 2 l^2} \lf -1  + \sqrt{1 + 8M G_5 \, l^2} \rt \;
. \label{r+}  \eeq  The substitution $\tau = i t$  makes the metric
positive definite and, by the usual removal of the conical singularity
at $r_+$, yields a periodicity in $\tau$ of  \beq  \beta = {2 \pi r_+
\over 1 + 2r_+^2 l^2 } \; ,\label{invtemp} \eeq  which is identified
with the inverse temperature of the black hole. The entropy is given
by   \beq  S = {A \over 4 G_5} = {\pi^2 r_+^3 \over 2 G_5} \; ,
\label{S}   \eeq  where $A$ is the ``area'' (that is 3-volume) of the
horizon.

We shall take the dual conformal field theory to be ${\cal N} = 4$,
U(N) super-Yang-Mills theory.   But since it is only possible to do
calculations in the weak coupling regime, we shall consider only the
free field limit of Yang-Mills theory. Then, in the high-energy regime
which  dominates the state counting, the spectrum of free fields on a
sphere  is essentially that of blackbody radiation in flat space, with
$8N^2$ bosonic and $8N^2$ fermionic degrees of freedom. The entropy is
therefore \beq  S_{\rm CFT} = {2 \over 3} \pi^2 N^2 \, V_{\rm CFT} \,
T^3_{\rm CFT} \; . \label{ent}  \eeq We would like to evaluate this
``holographically", i.e. by substituting physical data taken from the
boundary of the black hole spacetime. At fixed $r \equiv r_0 \gg r_+$,
the boundary line element tends to  \beq  ds^2 \to r_0^2 \left [ - l^2
dt^2 + d \Omega_3^2 \right ] \; .  \eeq The physical temperature at
the boundary is consequently red-shifted to \beq  T_{\rm CFT} =
{T_{BH} \over \sqrt{-g_{tt}}} = {T_{BH} \over lr_0} \; , \eeq  while
the volume is \beq  V_{\rm CFT} = 2 \pi^2 r_0^3 \; .   \eeq To obtain
an expression for $N$, we invoke the AdS/CFT correspondence.
Originating in the near horizon geometry of the D3-brane solution in
IIB supergravity, the correspondence \cite{adscft}, relates $N$ to the
radius of $S^5$ and the cosmological constant: \beq  R^2_{S^5} =
\sqrt{4 \pi g_s {\a '}^2 N} = {1 \over l^2} \; .  \eeq  Then, since
\beq  (2 \pi)^7 g_s^2 {\a '}^4 = 16 \pi G_{10} =  16 {\pi^4 \over l^5}
G_5 \; , \eeq  we have \beq  N^2 = {\pi \over 2 l^3 G_5} \;
. \label{N} \eeq  Substituting the expressions for $N$, $V_{\rm CFT}$
and $T_{\rm CFT}$ into  Eq. (\ref{ent}), we obtain \beq S_{\rm CFT} =
{1 \over 12} {\pi^2 \over l^6 G_5} \lf {1 + 2r_+^2 l^2 \over r_+} \rt
^{\! \! 3}  \; , \eeq  which, in the limit $r_+ l \gg 1$, reduces to
\beq S_{\rm CFT}  = {2 \over 3} {\pi^2 r_+^3 \over G_5} = {4 \over 3}
S_{\rm BH}\; ,  \eeq  in agreement with the black hole result,
Eq. (\ref{S}), but for a numerical factor of 4/3.

Similarly, the red-shifted energy of the conformal field theory
matches the black hole mass, modulo a coefficient.  The mass above the
anti-de Sitter background is  \beq  M' = {3 \pi \over 4} M \; .  \eeq
This is the AdS equivalent of the ADM mass, or energy-at-infinity.
The corresponding expression in the field theory is \beq
U^{\infty}_{\rm CFT} = \sqrt{-g_{tt}}  {\pi^2 \over 2} N^2 \, V_{\rm
CFT} \, T^4_{\rm CFT} = {\pi \over 2} r_+^4 l^2 = {4 \over 3} M' \; ,
\label{U} \eeq  where $U^{\infty}_{\rm CFT}$ is the conformal field
theory energy red-shifted to infinity, and we have again taken the
$r_+ l \gg 1$ limit.  The $4/3$ discrepancy in Eqs. (\ref{S}) and
(\ref{U}) is construed to be an artifact of having calculated the
gauge theory entropy in the free field limit rather than in the strong
coupling limit required by the correspondence; intuitively, one
expects the free energy to decrease when the coupling increases. The
4/3 factor was first noticed in the context  of D3-brane
thermodynamics \cite{fourthirds}.  Our approach differs in that we
take the idea of holography at face value, by explicitly reading
physical data from the boundary of spacetime; nonetheless,
Eq. (\ref{N}) refers to an underlying brane solution.

At this level, the correspondence only goes through in the $r_+ l>>1$
limit. Note that, in terms of the conformal field theory $r_+ l =
T_{\rm CFT} r_0$. The limit we have taken means that $T_{\rm CFT} \gg
1 / r_0$, allowing us to neglect  finite-size effects in the field
theory which we have implicitly done in calculating the entropy in
(\ref{ent}).

\section{Five-Dimensional Rotating AdS Black Holes}

The general rotating black hole in five dimensions has two independent
angular momenta. Here we consider the case of a rotating black hole
with one angular momentum in an ambient AdS space.  The line element
is \cite{5dadskerr} \begin{eqnarray} ds^2 & = & - {\D \over \r^2} \lf
dt - {a \sin^2 \th \over \Xi _a} d \phi  \rt ^{\! \! 2}  \nonumber \\
& + & {\D_\th \sin^2 \th \over \r^2} \lf a \, dt - {\lf r^2 + a^2 \rt
\over \Xi} d\phi \rt ^{\! \! 2} \nonumber \\ & & + {\r^2 \over \D}
dr^2 + {\r^2 \over \D_\th} \, d \th^2  + r^2 \cos^2 \th \, d \psi ^2
\; , \label{ds2} \end{eqnarray} where $0 \leq \phi , \psi \leq 2 \pi$
and $0 \leq \th \leq \pi / 2$, and \begin{eqnarray} \D & = & \lf r^2 +
a^2 \rt \lf 1 + r^2 l^2 \rt - 2MG_5 \nonumber \\ \D_\th & = & 1 - a^2
l^2 \cos^2 \th \nonumber \\ \r^2 & = & r^2 + a^2 \cos^2 \th \nonumber
\\ \Xi & = & 1 - a^2 l^2 \; .  \end{eqnarray} This solution is an
anti-de Sitter space with curvature given by  \beq  R_{ab} = - 4 l^2
g_{ab} \; .  \eeq  The horizon is at  \beq  r^2_+ = {1 \over 2 l^2}
\lf - (1 + a^2 l^2) +  \sqrt{(1 + a^2 l^2)^2 + 8 M G_5 \, l^2} \rt \;
, \eeq which can be inverted to give \beq  M G_5 = {1 \over 2}(r_+^2 +
a^2)(1+r_+^2l^2) \; .  \eeq The entropy is one-fourth the ``area'' of
the horizon: \beq  S = {1 \over 2 G_5} {\pi^2 \lf r_+^2 +a^2\rt r_+
\over \lf 1-a^2l^2\rt} \; . \label{SBH}  \eeq  The entropy diverges in
two different limits: $r_+ \to \infty$ and $a^2 l^2 \to 1$. The first
of these descibes an infinite temperature and infinite radius black
hole, while the second corresponds to  ``critical angular velocity'',
at which the Einstein universe at infinity has to rotate at the speed
of light. The inverse Hawking temperature is  \beq  \beta = {2 \pi \lf
r_+^2 + a^2 \rt \over r_+ \lf 1 + a^2 l^2 + 2 r_+^2 l^2 \rt} \;
. \label{beta}  \eeq  The mass above the anti-de Sitter background is
now  \beq  M' = {3 \pi \over 4 \Xi} M  \; , \label{mass}  \eeq the
angular velocity at the horizon is \beq \Omega_H = {a \Xi \over r_+^2
+ a^2} \; , \eeq and the angular momentum is defined as \beq
J_{\phi}=  {1 \over 16 \pi} \int_S \, \epsilon_{abcde} \del^{d}
\psi^{e} dS^{abc} = {\pi M a \over {2 \Xi^2}} \; , \label{jbh} \eeq
where $\psi^a= \lf {\pl \over \pl \phi} \rt ^{\! a}$  is the Killing
vector conjugate to the angular momentum in the $\phi$ direction, and
S is the boundary of a hypersurface normal to $\lf {\pl \over \pl t}
\rt ^{\! a}$, with $dS^{abc}$ being the volume element on S.

Following methods discussed in \cite{hawkingpage,thermalphase}, one
can derive a finite action for this solution from the regularized
spacetime  volume after an appropriate matching of hypersurfaces at
large $r$.  The result is \beq  I = {\pi^2 \lf r_+^2 + a^2 \rt ^{\! \!
2} (1 - r_+^2 l^2) \over 4 G_5 \Xi r_+ (1 + a^2 l^2 + 2 r_+^2 l^2 )}
\; . \label{f}  \eeq As noted in \cite{thermalphase},  the action
changes sign at $r_+ l = 1$, signalling the presence of a phase
transition in the conformal field theory. For $r_+ l > 1$, the theory
is in an unconfined phase and has a free energy proportional to
$N^2$. One can also check that the action satisfies the thermodynamic
relation  \beq   S = \beta(M' - J_{\phi} \Omega _H ) - I \; .  \eeq
It is interesting to note that, by formally dividing both the free
energy,  $F = I / \beta$, and the mass by an arbitrary volume, one
obtains an equation of state:  \beq  p = {1 \over 3} {r_+^2 l^2 - 1
\over r_+^2 l^2 + 1} \, \rho \; , \label{es}  \eeq  where $p = - F/V$
is the pressure, and $\rho$ is the energy density.  In the limit $r_+
l \gg 1$ that we have been taking, this equation becomes  \beq   p =
{1 \over 3} \rho  \; ,  \eeq  as is appropriate for the equation of
state of a conformal theory. This suggests that if a conformal field
theory is to reproduce the thermodynamic properties of this
gravitational solution, it has to be in such a limit.

\section{The dual CFT description}

The gauge theory dual to supergravity on $AdS_5 \times S^5$ is ${\cal
N} = 4$ super Yang-Mills with gauge group $U(N)$ where $N$ tends to
infinity \cite{adscft}.  The action is  \begin{eqnarray}  S = \int d^4
x \rg ~ {\rm Tr}  \big(  -{1 \over 4 g^2} F^2 + {1 \over 2} \lf D \Phi
\rt ^2 \nonumber \\ + {1 \over 12} R \Phi^2 + \bar{\psi} \Slash D \psi
\big) \; .  \end{eqnarray}  All fields take values in the adjoint
representation of U(N). The six scalars, $\Phi$, transform under
$SO(6)$ R-symmetry, while the four Weyl fermions, $\psi$, transform
under $SU(4)$, the spin cover of $SO(6)$.  The scalars are conformally
coupled; otherwise, all fields are massless. We shall again take the
free field limit.  The angular momentum operators can be computed from
the relevant components of the stress energy tensor in spherical
coordinates.  This approach is to be contrasted  with
\cite{steve,steve2,steve3,finnmirjam}  in which generators of
R-rotations are used corresponding to spinning D-branes.

The free energy of the gauge theory is given by \begin{eqnarray}
F_{\rm CFT} = +T_{\rm CFT} \sum_i \eta_i \int_0^{\infty} dl_i  \int d
m^{\phi}_i \int d m^{\psi}_i  \nonumber \\ \ln \lf 1 - \eta_i e^{-
\beta (\omega _i - m^{\phi}_i \Omega_{\phi})} \rt \; , \label{free}
\end{eqnarray}  where $i$ labels the particle species, $\eta = + 1$
for bosons and -1 for fermions, $l_i$ is the quantum number associated
with the total orbital angular momentum of the ith particle, and
$m^{\phi ( \psi )}_i$ is its angular momentum component in the  $\phi
(\psi)$ direction. Here $\Omega$ plays the role of a ``voltage'' while
the ``chemical potential'' $m^{\phi} \Omega$ serves to constrain the
total angular momentum of the system.

The free energy is easiest to evaluate in a corotating frame, which
corresponds to the constant-time foliation choice of  hypersurfaces
orthogonal to $t^a$. Since, at constant $r = r_0$, the boundary has
the metric \begin{eqnarray} ds^2 & = & r_0^2 \big[ - l^2 dt^2 + {2a
l^2 \sin^2 \th \over \Xi} dt \, d\phi + {\sin^2 \th \over \Xi} d
\phi^2 \nonumber \\ &+& {d\th^2 \over \D_\th} + \cos^2 \th \, d\psi^2
\big] \; , \end{eqnarray} the constant-time slices of the corotating
frame have a spatial volume of \beq  V =  {2 \pi^2 r_0^3 \over{
1-a^2l^2}}\; . \label{volumecft} \eeq The spectrum of a conformally
coupled field on $S^3$ is essentially given by \beq \omega _l \sim {l
\over r_0} \; , \eeq where $l$ is the quantum number for total orbital
angular momentum.  Eq. (\ref{free}) can now be evaluated by making use
of the identities \begin{eqnarray} \int_0^{\infty} dx \, x^n \ln \lf 1
- e^{-x + c} \rt  & = & - \Gamma(n+1) {\rm Li}_{n+2} (e^c) \nonumber
\\ & = & - \Gamma(n+1) \sum_{k = 1}^{k= \infty}  {e^{kc} \over
k^{n+2}} \nonumber \; ,  \\ \int dx \, x \, {\rm Li}_2(e^{-ax+c}) &=&
- {1 \over a^2} \big[ ax \, {\rm Li}_3(e^{-ax+c}) \nonumber \\ & + &
{\rm Li}_4(e^{-ax+c}) \big] \; , \end{eqnarray} where ${\rm Li}_n$ is
the nth polylogarithmic function, defined by the sum above. The result
is \beq  F_{\rm CFT} = -{\pi^4 \over 24} {r_0 \over {1 \over r_0^2} -
\Omega^2} \, (8 N^2) \, {T^4_{\rm CFT}} \; , \eeq yielding an entropy
of  \beq S_{\rm CFT} = {2 \over 3} {\pi^5 \over l^3 G_5} {r_0^3 \over
1 - \Omega ^2 r_0^2} \, T_{\rm CFT}^3 \; .	\label{entropycft}
\eeq  The physical temperature that enters the conformal field theory
is \beq  T_{\rm CFT} = {1 \over l r_0} T_{\rm BH} \; . \label{tempcft}
\eeq  Similarly, the angular velocity is scaled to  \beq  \Omega_{\rm
CFT} = {al^2 \over l r_0} \; . \label{omegacft}  \eeq  Substituting
Eqs. (\ref{tempcft}) and (\ref{omegacft}) into Eq. (\ref{entropycft})
and taking the high temperature limit as before, we have \beq  S_{\rm
CFT} = {2 \over 3 G_5} {\pi^2 r_+^3 \over (1 - a^2 l^2)}  = {4 \over
3} S_{\rm BH} \; .   \eeq  The inclusion of rotation evidently does
not affect the ratio of the black hole and field theory entropies.

In the corotating frame, the free energy is simply of the form $N^2 V
T^4$, with the volume given by Eq. (\ref{volumecft}). However, with
respect to a nonrotating AdS space, the free energy takes a more
complicated form since now the volume is simply $2 \pi^2 r_0^3$. By
keeping this volume and the temperature fixed, one may calculate the
angular momentum of the system with respect to the nonrotating
background: \beq  J^{\rm CFT}_{\phi} = \left. - {\pl F \over \pl
\Omega} \right |_ {V,T_{\rm CFT}}  = { a r_+^4 \pi \lf 1 + a^2 l^2
+2r_+^2 l^2 \rt ^4 \over  48 l^6 \Xi^2 \lf r_+^2 + a^2 \rt ^4} \; .
\eeq In the usual $r_+ l \gg 1$ limit, we obtain \beq   J_{\phi}^{\rm
CFT} = {2 \pi M a \over 3 \Xi^2} = {4 \over 3} J_{\phi}^{\rm BH} \; ,
\eeq so that the gauge theory angular momentum is proportional to the
black hole angular momentum, Eq. (\ref{jbh}), with a factor of 4/3.

The black hole mass formula, Eq. (\ref{mass}), refers to the energy
above the nonrotating anti-de Sitter background. We should therefore
compare  this quantity with the red-shifted energy in the conformal
field theory.  Here a slight subtlety enters.  Since the statistical
mechanical calculation gives the energy in the corotating frame, we
must add the center-of-mass rotational energy before comparing with
the black hole mass. Then we find that \beq U^{\infty}_{\rm CFT} =
\sqrt{-g_{tt}} \lf U_{\rm corotating} + J_{\rm CFT}  \Omega_{\rm CFT}
\rt = {4 \over 3} M' \; , \eeq with $M'$ given by Eq. (\ref{mass}),
evaluated at high temperature. Using $U^{\infty}_{\rm CFT} =
\sqrt{-g_{tt}} U_{\rm CFT}$ and previous expressions for thermodynamic
quantities, one may check that the first law  of thermodynamics is
satisfied.

\section{Discussion}

There are several interesting aspects of these results.  The first is
that the same relative factor that appears in the entropy appears in
the angular momentum and the energy. A priori, one has no reason to
believe that the functional form of the free energy will be such as to
guarantee this result (see, for example,\cite{steve}).  The second is
that the relative factor of 4/3 in the entropy is  unaffected by
rotation. Indeed, one could expand the entropy of the rotating system
in powers and inverse powers of the 't Hooft coupling. The
correspondence implies that  \beq  S_{\rm CFT} = \sum_m a_m \lm^m =
\sum_n b_n \lf {1 \over \sqrt{\lm}} \rt ^{\! \! n} = S_{\rm BH} \; .
\eeq  We may approximate the series on the gauge theory side as $a_0$
and on the gravity side as $b_0$. Then, generically, we would expect
these coefficients to be functions of the dimensionless rotational
parameter $\Xi$ so that $a_0 (\Xi) = f(\Xi) b_0(\Xi)$ with $f(\Xi = 1)
= 4/3$. Our somewhat unexpected result is that $f(\Xi) = 4/3$ has, in
fact, no dependence on $\Xi$. Ofcourse, in some sense the most
fascinating aspect is that the ratio between strong and weak coupling
is such a simple rational number. A futher understanding of this
property of SYM would be very desirable.

Similar properties have been investigated with the incorporation of
finite size effects by \cite{real,lopez}.

\section{Acknowledgments}

We would like to thank Jos\'e Barb\'on, Roberto Emparan, and  Kostas
Skenderis for helpful discussions. D. B. is supported by European
Commission TMR programme ERBFMRX-CT96-0045.  M. P. is supported by the
Netherlands Organization for Scientific Research (NWO).

\end{document}